\documentclass[aps,prd,reprint,groupedaddress,showpacs,amsmath,nofootinbib]{revtex4-1}
\usepackage{hyperref}
\usepackage{graphicx}

\begin{document}
\title{Gaussian Approximation of Peak Values in the Integrated Sachs-Wolfe Effect}

\author{Simone Aiola}
\email[Email: ]{sia21@pitt.edu}
\author{Arthur Kosowsky}
\email[Email: ]{kosowsky@pitt.edu}
\author{Bingjie Wang}
\affiliation{Department of Physics and Astronomy, University of Pittsburgh, Pittsburgh, PA 15260 USA}
\affiliation{Pittsburgh Particle Physics, Astrophysics, and Cosmology Center (PITT-PACC), Pittsburgh PA 15260}
\date{\today}

\begin{abstract}
The accelerating expansion of the universe at recent epochs is encoded in the cosmic microwave background: 
a few percent of the total temperature fluctuations are generated by evolving gravitational potentials which trace the large-scale structures in the universe. 
This signature of dark energy, the Integrated Sachs-Wolfe Effect, has been detected by averaging temperatures in the WMAP sky maps corresponding 
to the directions of superstructures in the Sloan Digital Sky Survey data release 6. 
We model the maximum average peak signal expected in the standard $\Lambda$CDM cosmological model, using Gaussian random realizations of the microwave sky, 
including correlations between different physical contributions to the temperature fluctuations and between different redshift ranges of the evolving gravitational potentials. 
We find good agreement with the mean temperature peak amplitude from previous theoretical estimates based on large-scale structure simulations, but with larger statistical uncertainties. 
We apply our simulation pipeline to four different foreground-cleaned microwave temperature maps from Planck and WMAP data, finding a mean temperature peak signal
at previously identified sky locations which exceeds our theoretical mean signal at a statistical significance of about $2.5\sigma$ and which differs from a null signal at $3.5\sigma$.
\end{abstract}


\maketitle
\section{Introduction\label{secI}}
The current state of accelerated expansion of the universe has been well established through a combination of the type-Ia supernova Hubble diagram \cite{Riess,Perlmutter}, primary and lensing-induced anisotropies in the cosmic microwave background (CMB) \cite{Sherwin,VanEngelen,PLANCK_PARAMS}, and measurements of baryon acoustic oscillations \cite{BAO}. 
Such an expansion, believed to be driven by dark energy, leaves an imprint in the large-scale cosmic structure (at redshifts in a range of $z\lesssim 2$), as well as on the CMB temperature fluctuations.
Gravitational potentials evolve in time due to the accelerating expansion, giving a net change in energy to photons traversing an underdense or overdense region. This effect, known as the \emph{late-time Integrated Sachs-Wolfe effect} (late-ISW) \cite{SachsWolfe}, is described by the following integral along the line-of-sight:
	\begin{equation}
		\Theta(\mathbf{\hat{n}}) \equiv \frac{\Delta T}{T_{0}}=-2\int_0^{\chi^{\star}} d\chi g(\tau)\dot{\Phi}(\chi\hat{\mathbf{n}},\eta_0-\chi)
	\end{equation} 
where $g(\tau)=e^{-\tau(\eta_0-\chi)}$ is the visibility function as a function of the optical depth $\tau$, the derivative of the Newtonian gravitational potential $\Phi$ is with respect to the conformal time, $\eta_{0}$ is the present value of the conformal time, $\chi^{\star}$ is the comoving distance to the surface of last scattering, and $T_0$ is the isotropic CMB blackbody temperature, corresponding to the multipole $\ell=0$. The late-ISW effect creates temperature anisotropies mostly on relatively large angular scales ($\theta > 3^{\circ}$).
A detection of this signal in a spatially flat universe represents an independent test for dark energy  \cite{Kamionkowski}, and in principle a useful tool to characterize its properties and dynamics \cite{Giannantonio08}.

In $\Lambda$CDM cosmological models, this secondary CMB anisotropy contributes only around $3\%$ of the total variance of the temperature sky, while having a Gaussian random distribution to a very good approximation, and hence it cannot be detected from temperature data alone. Nevertheless, it is strongly correlated with the large-scale galaxy distribution \cite{Crittenden}, and recently the angular cross-power spectrum $C_{\ell}^{Tg}$ between CMB temperature and galaxies has been exploited to detect the late-ISW signal \cite{Fosalba, Francis, Afshordi, Boughn, Nolta, Padmanabhan, Rassat, Sawangwit, Vielva, Cabre, Ho, Planck_ISW, Monteagudo13, Kovacs, Ferraro} (see Table 1 of \cite{Nishizawa} for a detailed list of related works).
A similar correlation was detected in pixel space, corresponding to the presence of hot and cold spots in the CMB sky preferentially centered on superstructures (\cite{Granett}, GNS08 hereafter). This strong detection exploited a novel technique involving photometric analysis of stacked CMB patches from the WMAP 5-year sky maps \cite{hinshaw09} centered on $100$ superstructures (50 biggest superclusters and 50 biggest supervoids) detected in the Sloan Digital Sky Survey (SDSS) Data Release 6 \cite{SDSS6}, covering a sky area of $7500$ square degrees in a redshift range $0.4<z<0.75$. In this redshift range, the expected cross-correlation spectrum peaks at $\ell \simeq 20$ ($\theta \simeq 4^{\circ}$) (\cite{Monteagudo}, HMS13 hereafter), which motivated the use of a compensated top-hat filter of $4^{\circ}$ radius to enhance the signal 
\cite{GNS08Sup}. The mean temperature fluctuation reported by GNS08 of $\overline{T} = 9.6$ $\mu$K shows a departure from the null signal at a significance  of $4.4\sigma$.  Recently, the Planck satellite collaboration has confirmed the detection of the late-ISW effect with a statistical significance ranging from $2.5\sigma$ to $4.0\sigma$ (depending on the method involved) (\cite{Planck_ISW}, PLK13 hereafter). The strongest Planck detection is associated with the stacking analysis, using the GNS08 catalog, 
giving an average peak amplitude of $\overline{T} = 8.7 \mu K$, which is consistent with the value found by GNS08 using the WMAP temperature map.

As pointed out by \cite{Ferraro}, the temperature-galaxy cross correlation requires prior knowledge of the galaxy bias, which may dominate the detection significance
and consistency tests of the underlying cosmological model. In contrast, the technique of stacking on the largest superstructures in a large-scale structure survey
does not rely on any knowledge regarding the galaxy bias, apart from the fact that visible matter traces dark matter. In addition, the GNS08 technique is based on an extreme-value statistic: in principle, it is sensitive to small departures from the $\Lambda$CDM model which may not significantly affect the cross-correlation $C^{Tg}_{\ell}$. 
On the other hand, substantial control over systematic errors is required to carry out such an analysis.

It has been argued that the strong signal detected by GNS08 is in tension with the underlying $\Lambda$CDM model \cite{Nadathur,Flender}.
Analytical estimates of the stacked late-ISW signal in a comoving volume that corresponds to that probed by GNS08 predict an average signal of $\overline{T} = 2.27 \pm 0.14 \mu K$(\cite{Flender}, FHN13 hereafter), where the reported error is due to cosmic variance. The same work confirms this estimate using
late-ISW maps constructed from N-body simulations which include the second-order Rees-Sciama contribution \cite{ReesSciama}.The discrepancy with the GNS08 measurement has a significance greater than 3$\sigma$. Other cosmological models have been considered to explain the discrepancy, including primordial non-Gaussianities \cite{Monteagudo} and $f(R)$ gravity theories \cite{Cai}, but neither seems adequate to explain the strong detected signal.

A less interesting but more plausible possibility is that the strong detected signal is the result of correlations of the late ISW signal with other sources of temperature anisotropy, which may boost the mean temperature of the identified top-ranked peaks.
The current theoretical predictions of the stacked late-ISW signal do not include correlations between ISW temperature fluctuations formed at different redshifts. In HMS13, the primary temperature fluctuations, formed at redshift $z\simeq 1100$, were considered uncorrelated with the secondary anisotropies and simply added to gaussian random generated late-ISW maps. These high-redshift fluctuations are partially correlated with the secondary temperature anisotropies, at a level that depends on the underlying cosmological model. More importantly, we expect a non-negligible correlation between the late-ISW signal, traced by superstructures in GNS08 in the redshift range $0.4 < z < 0.75$, and the late-ISW effect due to structures at either higher or lower redshift. 

In this work, we provide a complete description of these correlations through simulated skies based on simple linear perturbation theory. Temperature fluctuations on large scales
result from gravitational potential perturbations in the linear regime (see \cite{Chen} for alternative proposal). If the primordial perturbations are a Gaussian random field, which appears to be an excellent
approximation to the observed large-scale structure \cite{Papai10}, the statistical properties of the CMB sky on large angular scales are completely specified by the temperature power spectrum $C_{\ell}^{TT}$. We generate Gaussian random realizations of the CMB sky using the linear power spectra for its various physical components, including correlations between them. This is an easy computational process, in contrast to extracting large-angle late-ISW maps from large-box N-body cosmological simulations \cite{Watson14, Hotchkiss}.  The approach we adopt 
in this paper allows full characterization of cosmic variance with a random sample of simulated skies, and it automatically accounts for the effects of the largest-scale perturbation modes beyond the reach of N-body simulations. We then reanalyze foreground-cleaned CMB temperature maps, processed to match the procedure adopted in our sky simulations. This last step guarantees that the discrepancy between theoretical estimates and the measured signal is not due to different analysis procedures.
Our simulated late-ISW mean peak temperature signal is consistent with previous estimates, but with a wider spread of values. Correlations between temperature signals increase the expected mean value as well as the spread slightly. The main reason for this larger spread, however, is the noise associated to the uncorrelated fluctuations at scales of our interest, and thus reduces the statistical significance of the discrepancy between theory and experiment to around $2.5\sigma$ when compared with our measured values from CMB maps.

We present our work as follows. In Section \hyperref[secII]{\ref*{secII}}, we describe an algorithm to generate realistic temperature maps, including spatial filtering and all correlations between temperature components. We then present the pipeline of our simulations in Section \hyperref[secIII]{\ref*{secIII}}, and the resulting distribution of late-ISW mean peak temperatures. In Section \hyperref[secIV]{\ref*{secIV}}, we apply the same procedure to the Planck CMB temperature maps. Finally, Section \hyperref[secV]{\ref*{secV}} concludes with a discussion of possible sources of systematic errors, a comparison with other late-ISW detection techniques, and future prospects for resolving the discrepancy between theory and measurements with wider and deeper large-scale structure surveys. 

\section{Correlated Components of the Temperature Sky\label{secII}} 

\begin{figure}[t!]
	\centering
		\includegraphics[scale=0.6]{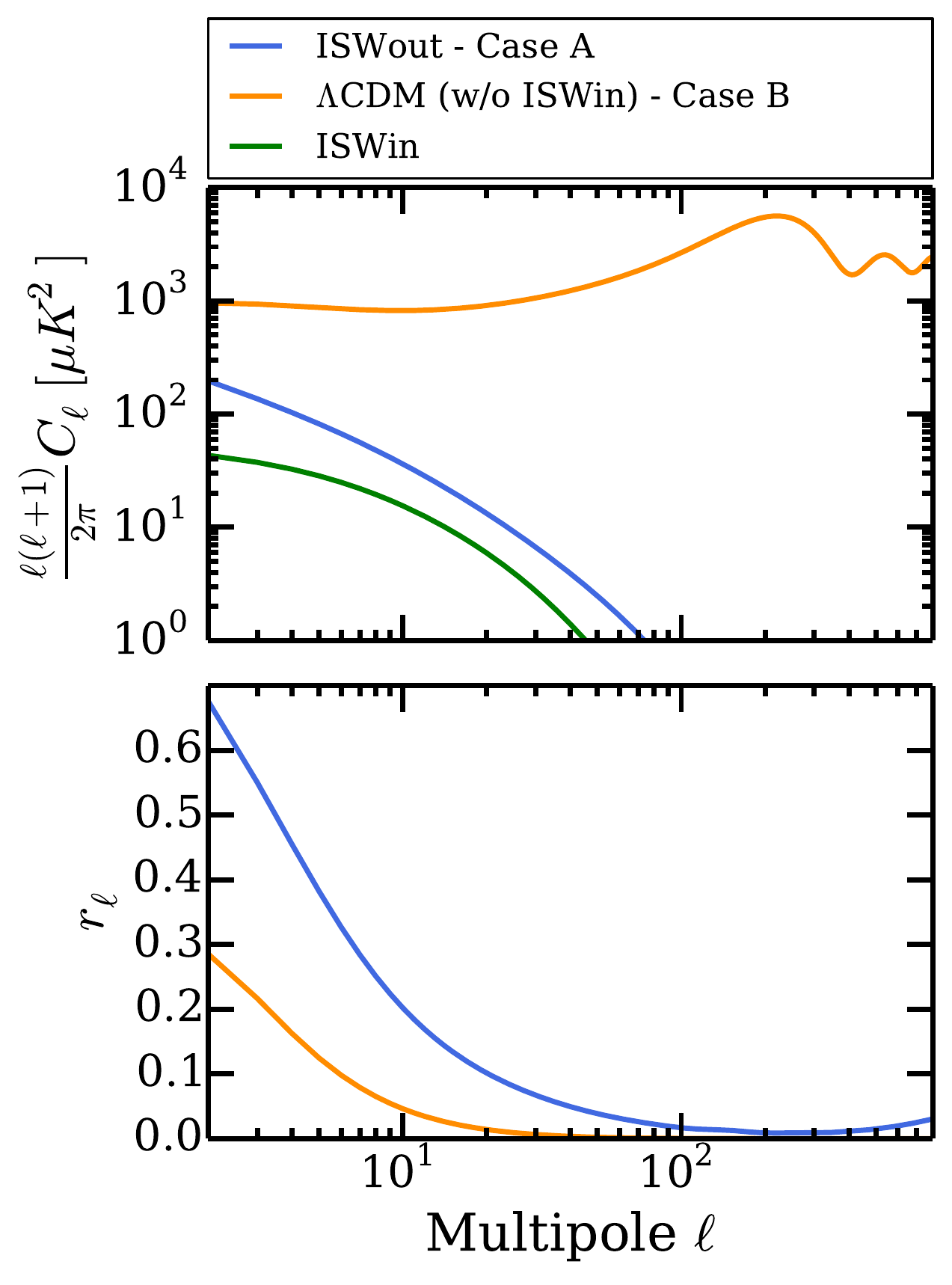}
		\caption{Top: Angular power spectra in $\Lambda$CDM, for the ISW effect due to structure in the redshift range $0.4 < z < 0.75$ (``ISW--in'', green), ISW effect outside of this redshift range (``ISW--out'', blue), and all temperature perturbation components except for ISW--in (yellow). Bottom: Correlation coefficients between ISW--in
	and ISW--out (blue), and between ISW--in and all other temperature perturbation components (yellow).} 
	\label{fig:cls}
\end{figure}

The $\Lambda$CDM model is a compelling theory to describe the statistical properties of the CMB fluctuations, making precise predictions for the temperature power spectrum $C_{\ell}^{TT}$ \cite{WMAP9, PLANCK_PS}. 
Different physical processes contribute to the temperature fluctuations over a wide range of angular scales; the CMB temperature sky is well approximated by the sum of correlated
gaussian random fields, one for each physical component, such that
\begin{equation}
	\begin{aligned}
		\langle a_{\ell m}^{i}, a_{\ell ' m'}^{i \star} \rangle &= \delta_{\ell \ell '} \delta_{mm'} C_{\ell}^{ii}\\
		\langle a_{\ell m}^{i}, a_{\ell ' m'}^{j \star} \rangle &= \delta_{\ell \ell '} \delta_{mm'} C_{\ell}^{ij}\\
	\end{aligned}
	\label{cov_mat}
\end{equation}
where $i$ and $j$ are the components making up the observed temperature field $\Theta(\mathbf{\hat{n}})=\sum_i \sum_{\ell m}a_{\ell m}^{i}Y_{\ell m}$ and the power spectra satisfy the condition $C_{\ell}^{i i} C_{\ell}^{j j} \ge \big(C_{\ell}^{i j}\big)^2$ \cite{KKS}. This set of power spectra specify the covariance matrix of the temperature given a cosmological model. For the purposes of this work, we consider a 2-component sky described by a symmetric 2x2 covariance matrix. The first component, $C_{\ell}^{1,1}$, is always the late-ISW component of the temperature field, corresponding to the GNS08 redshift range (ISW--in, hereafter). For the second component, $C_{\ell}^{2,2}$, we consider two distinct cases:
\begin{itemize}
\item \textbf{Case A}: \emph{only} late-ISW generated outside the probed redshift range, corresponding to $0<z<0.4$ and $0.75<z<10$ (ISW--out, hereafter);
\item \textbf{Case B}: primary and secondary anisotropies generated outside the probed redshift range. Specifically, we consider the sum of ISW--out, early ISW after
recombination, and Sachs-Wolfe, intrinsic and Doppler contributions at last scattering.
\end{itemize}

The off-diagonal terms $C_{\ell}^{1,2}$ are calculated according to the specific case we consider. 
For a spatially flat, $\Lambda$CDM cosmological model with the best-fit Planck+WP+HighL parameters \cite{PLANCK_PARAMS}
we compute the covariance matrix in Eq.~(\ref{cov_mat}) with the numerical Boltzmann code \verb|CLASS v2.2|\footnote{\url{http://class-code.net/}} \cite{CLASS},
including the nonlinear effects calculated with \verb|Halofit| \cite{HALOFIT}. 
The correlated harmonic coefficients are generated via Cholesky decomposition as
\begin{equation}
	\begin{aligned}
		a_{\ell m}^{i}&= \sum_{k=1}^2 A_{\ell, ik}\zeta_k\\
		a_{\ell m}^{T} &= a_{\ell m}^{1} + a_{\ell m}^{2}\\
	\end{aligned}
	\label{alms}
\end{equation}
where $\zeta_k$ is a column vector composed of $2$ complex gaussian random numbers with zero mean and unit variance, and $A_{\ell}$ is a lower-diagonal real matrix which satisfies $C_{\ell}=A_{\ell}^TA_{\ell}$.
The $a_{\ell m}^{1}$ are the harmonic coefficients corresponding to the ISW--in component alone.

In Fig.~\ref{fig:cls}, we plot the unfiltered covariance matrix components as function of the multipole $\ell$. The top panel shows the diagonal terms. Note that the signal of interest, ISW--in, has a lower amplitude compared than the other components at all multipoles. Thus, the statistics of temperature peaks for an unfiltered map are completely dominated by the anisotropies generated at last scattering. A wise choice for an $\ell$-space filter is required (see below, Sec. \hyperref[secIII]{\ref*{secIII}}). 
The bottom panel shows the off-diagonal terms; we plot the normalized correlation coefficient
\begin{equation}
	r_{\ell} \equiv \frac{C^{i j}_{\ell}}{\sqrt{C^{i i}_{\ell} C^{j j}_{\ell}}}
\end{equation}
which satisfies the condition $|r_{\ell}| \leq 1$.
The correlation matrix cannot be considered diagonal, especially at low $\ell$ values.
In principle we expect a negative cross-correlation on large scales (i.e. $r_{\ell}<0$) due to the Sachs-Wolfe component: if we consider the entire late-ISW contribution (i.e., $0<z<10$), the cross-spectrum is dominated by the ISW-SW term, which gives an overall anti-correlation. In the case of interest (where we consider shells of late-ISW signal), the dominant part is the correlation between ISW--in and ISW--out. 
Notice that $r^{\small{\rm CaseA}}_{\ell}/r^{\rm CaseB}_{\ell} \simeq \sqrt{C^{2,2 (\rm CaseB)}_{\ell}/C^{2,2 (\rm CaseA)}_{\ell}}$, which implies that the mean value of the stacked signal is mainly enhanced by the ISW--out component. This peculiar effect is attributed to the wide range of $k-$modes, which couples the fluctuations of neighboring redshift regions. On the other hand, the mildly correlated primary fluctuations dominate the statistical error in averaged peak values.  Analytical signal and error estimates are possible but not simple \cite{PEAK_AKW}, so we compute both numerically in the following Section.

\section{Methodology and Analysis\label{secIII}}

The multipole region of our interest is dominated by cosmic variance. This problem is difficult to characterize using N-body simulations, so we generate random temperature maps from the power spectra and correlations to construct the statistical distribution of ISW mean peak amplitudes. The procedure described in this section is based on the FHN13 analysis, adapted to multi-component correlated sky maps.

\subsection{Harmonic-Space Filtering\label{secIII.A}}

To isolate the late-ISW peak signal in $\ell$-space, we apply the  $4^\circ$ compensated top-hat filter adopted by GNS08:
\begin{equation}
	F(\theta) =
  	\begin{cases}
   	(2\pi(1-\cos{\theta_F}))^{-1}, & 0<\theta<\theta_F, \\
   	\\
   	-(2\pi(\cos{\theta_F}-\cos{\sqrt{2}\theta_F}))^{-1},  & \theta_F<\theta<\sqrt{2}\theta_F,
  	\end{cases}
	\label{tophat_filter}
\end{equation} 
where $\theta_F=4^{\circ}$ is the characteristic filter radius. By performing a Legendre transform of the real-space filter $F(\theta) \rightarrow F_{\ell} = \int F(\theta)\mathcal{P}_{\ell}(\cos\theta) d\cos\theta$, we can compute a full-sky filtered map simply by rescaling the covariance matrix, $C_{\ell} \rightarrow C_{\ell}F^2_{\ell}B^2_{\ell}$, which also uses
an additional Gaussian beam smoothing $B_{\ell}$ with  \verb|FWHM|$=30'$ adopted by PLK13 to match the WMAP resolution. 
The compensated top-hat filter does not give a sharp cutoff in multipole space. However, it drops off faster than $\ell^{-2}$, which ensures the suppression of the small-scale fluctuations. 
At the scales enhanced by the filter $\ell \simeq 10-30$, the portion of the temperature fluctuations uncorrelated with the ISW--in signal  for Case B is approximately one order of magnitude larger than that for Case A, with a resulting increase in the scatter of the mean peak statistic.

\subsection{Simulation Pipeline\label{secIII.B}}

To identify the peaks of the late-ISW temperature fluctuations in the CMB sky map, GNS08 used the distribution of luminous red galaxies in \verb|SDSS DR6|
and looked for overdense and underdense regions. 
The top-ranked $100$ superstructures identified in the sample have a median radial length calculated at $z=0.5$ of $R_v \simeq 85 Mpc$ and $R_c \simeq 25 Mpc$ for voids and clusters respectively.
The corresponding normalized fluctuations of the gravitational potential are of the order $\Phi \simeq 10^{-4}$ \cite{GNS08Sup}. 
These gravitational potential fluctuations are still in the linear regime for standard structure growth.
\begin{figure}[t!]
	\centering
		\includegraphics[scale=0.34]{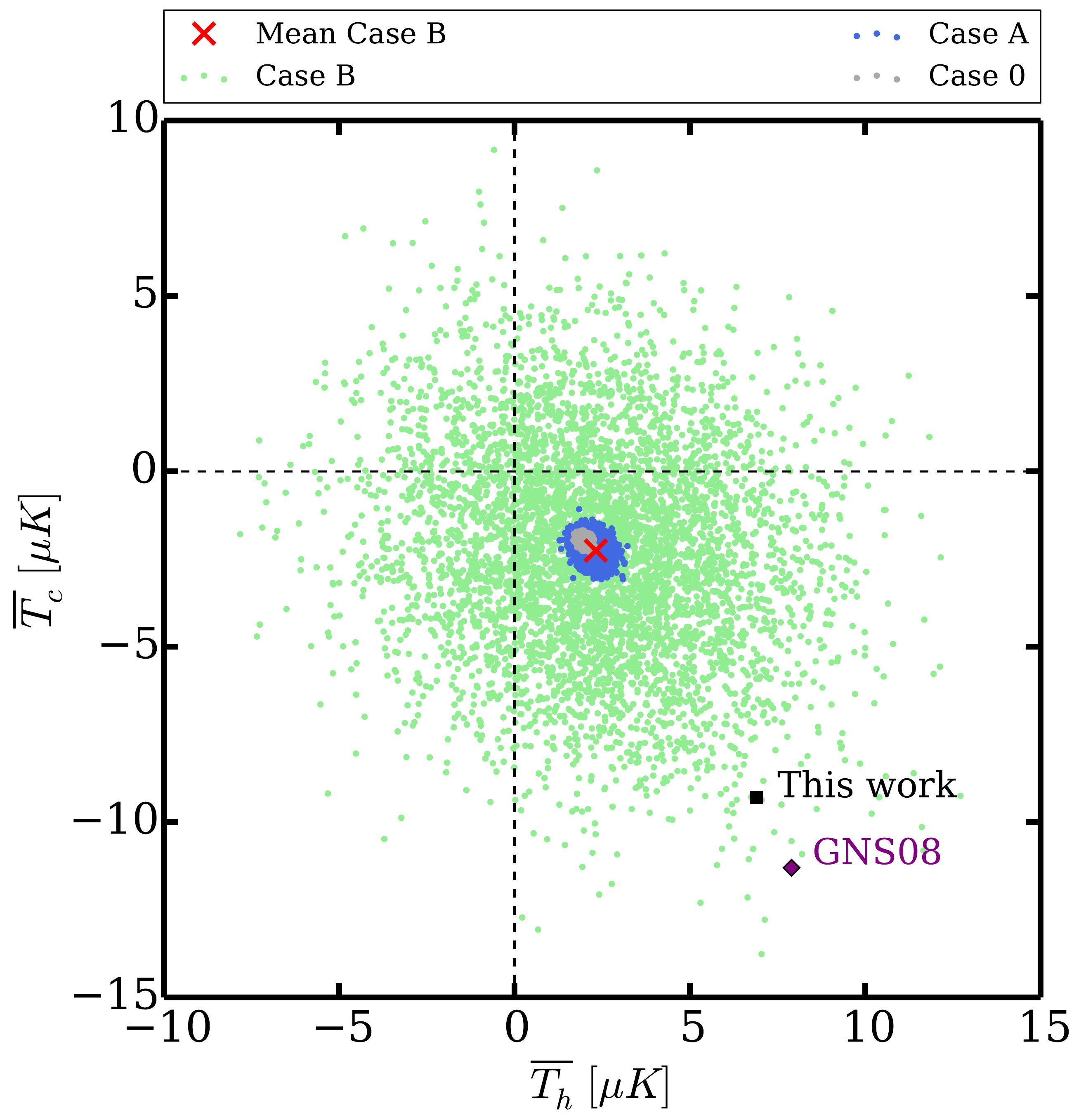}	
		\caption{The mean value of the filtered CMB temperature at the locations of the top 50 cold spots $\overline{T_{\rm cold}}$ and top 50 hot spots $\overline{T_{\rm hot}}$ of the ISW--in map component, corresponding to the late-ISW signal from structures in the redshift range $0.4 < z < 0.75$, for a sky fraction $f_{sky}=0.2$.
	Plotted are $(\overline{T_{\rm hot}},\overline{T_{\rm cold}})$ for 5000 randomly generated skies with all contributions to the CMB signal (green points). The red cross is at the location of the mean values
	of $\overline{T_{\rm cold}}$ and $\overline{T_{\rm hot}}$ for the 5000 model skies. For comparison, we plot  5000 model skies generated
	using only the ISW--in signal (gray points), and 5000 skies generated using the full late late-ISW signal but no other temperature components (blue points).
	Also displayed are the measured values from GNS08 (purple diamond) and from the analysis in Sec.~IV using Planck data (black square).}
	\label{fig:scatter_plot}
\end{figure}

Assuming perfect efficiency in detecting and ranking superstructures from large-scale structure distribution data, the observed GNS08 signal should match the theoretical expectation from averaging the CMB temperature fluctuations traced by the $100$ biggest fluctuations in the filtered late-ISW map over the redshift range of the survey \cite{Flender}.  
We generate correlated pairs of filtered random Gaussian maps, one for the ISW--in component and one for the other linear components of the temperature sky, using multipoles in both power spectrum $\ell \leq 800$; we use \verb|HEALPix|\footnote{\url{http://healpix.sf.net}} \cite{HEALPix} with \verb|NSIDE|=$256$. From the 
filtered ISW--in map, we identify the $50$ hottest maxima and $50$ coldest minima in a sky region of area $f_{\rm sky}=0.2$, corresponding to the sky fraction of the \verb|SDSS DR6| survey. Maxima and minima are identified pixel-by-pixel, testing whether or not the temperature of the central pixels is the greatest or the smallest of the $8$ surrounding pixels. Finally, we take the pixels corresponding to these extrema and average their values in the full sky map consisting of the sum of the two correlated random maps. We find
the average of the 50 hottest ISW--in maxima $\overline{T_h}$ and 50 coldest ISW--in minima $\overline{T_c}$ separately, and we also compute the combined mean value as 
$\overline{T_m} = (\overline{T_h} - \overline{T_c})/2$. For comparison, we also calculate the same quantities for the ISW--in map only, which we call \textbf{Case 0}. This procedure is performed on an ensemble of $5000$ random generated skies.

The procedure adopted here gives an upper bound on the theoretical signal from clusters and voids identified in any specific tracer of large-scale structure: we
simply assume that the 50 largest voids and 50 largest clusters in a sky region are correctly identified. Any error in identifying these features will lead to a smaller
mean signal. Since the measured signal is larger than the expected theoretical maximum signal, errors in cluster identification will increase the difference between
theory and measurement quantified in the next section.

\subsection{Results and Comparison with Previous Work} 

\begin{table}[t!] 
\caption{Results from Gaussian random skies, stacked on peaks of the ISW--in signal (the ISW generated for structure in the redshift range  $0.4<z<0.75$). 
The simulated skies are constructed from the angular power spectra in the standard $\Lambda$CDM cosmology, smoothed with a Gaussian beam of FWHM 30' and a compensated
top hat filter of radius $4^{\circ}$, Eq.~(\ref{tophat_filter}).
We report the mean and the standard deviation of the stacks on the locations of the 50 hottest ISW--in spots $\overline{T_h}$, 50 coldest ISW--in spots $\overline{T_c}$, and 
the mean magnitude for all 100 spots $\overline{T_m}$, calculated from $5000$ random realizations of the microwave sky, including correlations between the ISW--in signal and
other sky components. These values are presented for ISW--in skies only (Case 0), ISW--in plus ISW--out skies (Case A), and realistic skies including early ISW, intrinsic, and Doppler contributions to the sky temperature (Case B). The theoretical prediction from FHN13 and the measured value from GNS08 are reported for comparison.}
\centering  
\begin{tabular}{c | c | c | c}
\hline\hline 
&&&\\
Case & $\quad$ $\overline{T_h}$ $[\mu K]$ $\quad$ & $\quad$ $\overline{T_c}$ $[\mu K]$ $\quad$ & $\quad$ $\overline{T_m}$ $[\mu K]$\\ [0.5ex]
\hline\hline 
\textbf{Case 0}& $1.97 \pm 0.09$ &$-1.97 \pm 0.09$& $1.97 \pm 0.07$\\
\textbf{Case A} & $2.23 \pm 0.25$ &$-2.23 \pm 0.25$& $2.23 \pm 0.20$\\
\textbf{Case B} & $2.30 \pm 3.1$ &$-2.30 \pm 3.1$& $2.30 \pm 2.32$\\
\hline\hline
\textbf{FHN13}& - &-& $2.27 \pm 0.14$\\
\textbf{GNS08} & $7.9 \pm 3.1$ & $-11.3 \pm 3.1$ & $9.6 \pm 2.22$\\
\hline\hline
\end{tabular} 
\label{table:sims_res} 
\end{table} 
The results of our simulations are presented in Table~\ref{table:sims_res} and visually summarized in Fig.~\ref{fig:scatter_plot} and Fig.~\ref{fig:hist_plot}. 
As expected for random realizations of a Gaussian field, $|\overline{T_h}|=|\overline{T_c}|$. The mean peak signal for the full simulated
sky maps (Case B) is $2.30\pm 2.32$ $\mu$K, compared to the GNS08 measurement of $9.6$ $\mu$K, a discrepancy at
a significance of $3.1\sigma$. Our discrepancy is about the same size as previous analyses, but the significance is somewhat lower.
This is due to our inclusion of all components in the microwave temperature map and their correlations, which increases the uncertainty
in our predicted values. The central value of our ISW--in peak signal, $1.97$ $\mu$K (Case 0), is lower by 0.30 $\mu$K than the signal
predicted in FHN13, which is expected due to a difference in the underlying cosmological models used. However, the difference is
small compared to the statistical uncertainty for the full sky signal (Case B). The central value of our full-sky peak signal is also higher
than the ISW--in peak signal by $0.33$ $\mu$K; this difference is due to the correlations between the ISW--in signal and the other components
which are included in the Case B peak signal. 

\section{The Stacked ISW Signal Using Planck Sky Maps\label{secIV}}
\begin{figure}[t!]
	\centering
		\includegraphics[scale=0.32]{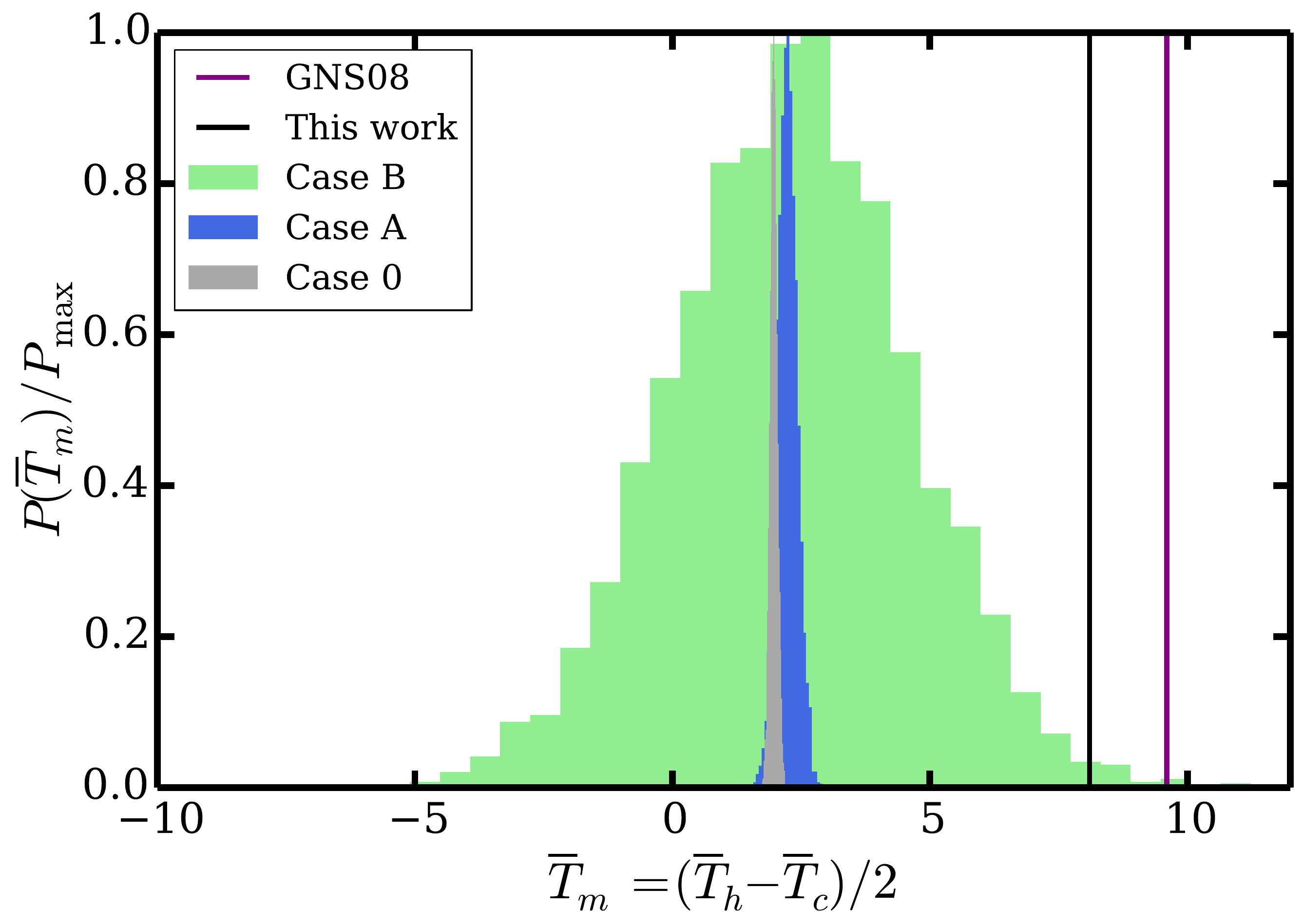}	
		\caption{The combined mean value of the filtered CMB temperature at the locations of the top 50 cold spots and top 50 hot spots of the ISW--in map component, corresponding to the late-ISW signal from structures in the redshift range $0.4 < z < 0.75$, for a sky fraction $f_{sky}=0.2$. Plotted are the distributions (normalized to the maximum value) of the combined mean temperature $(\overline{T_{\rm hot}} - \overline{T_{\rm cold}})/2$ obtained from 5000 simulated skies, for the three difference cases considered in this work. Also displayed are the measured values from GNS08 (purple vertical line) and from the analysis in Sec.~IV using Planck data (black vertical line).}
		\label{fig:hist_plot}
\end{figure}

The original late-ISW peak analysis in GNS08 used WMAP sky maps, and PLK13 confirmed the measured value using Planck data. 
Here we obtain the measured late-ISW signal from publicly available foreground-cleaned maps based on Planck and Planck+WMAP data, 
using the same sky locations as GNS08.
The purpose of this re-analysis is testing the significance of the discrepancy by using the same analysis pipeline as the simulations in Sec.~\ref{secIII}, to
ensure that the difference between the model and the measured value is not due to any inconsistency in how the data and simulations are treated.

We use four different foreground-cleaned CMB temperature maps, based on different component separation approaches. Two are public CMB temperature maps from the \verb|Planck| collaboration\footnote{\url{http://www.sciops.esa.int/wikiSI/planckpla}}, namely \verb|SMICA| and \verb|NILC|\cite{PLANCK_CS}. The other two maps are based on the \verb|LGMCA| method\footnote{\url{http://www.cosmostat.org/CosmoStat.html}} from the recent work in Ref.~ \cite{Bobin}. The \verb|PR1| map uses only \verb|Planck DR1| data \cite{PLANCK_GEN}, and the \verb|WPR1| map uses both \verb|Planck DR1| and \verb|WMAP9| data \cite{WMAP9}.

\begin{figure}[t!]
	\centering
		\includegraphics[scale=0.6]{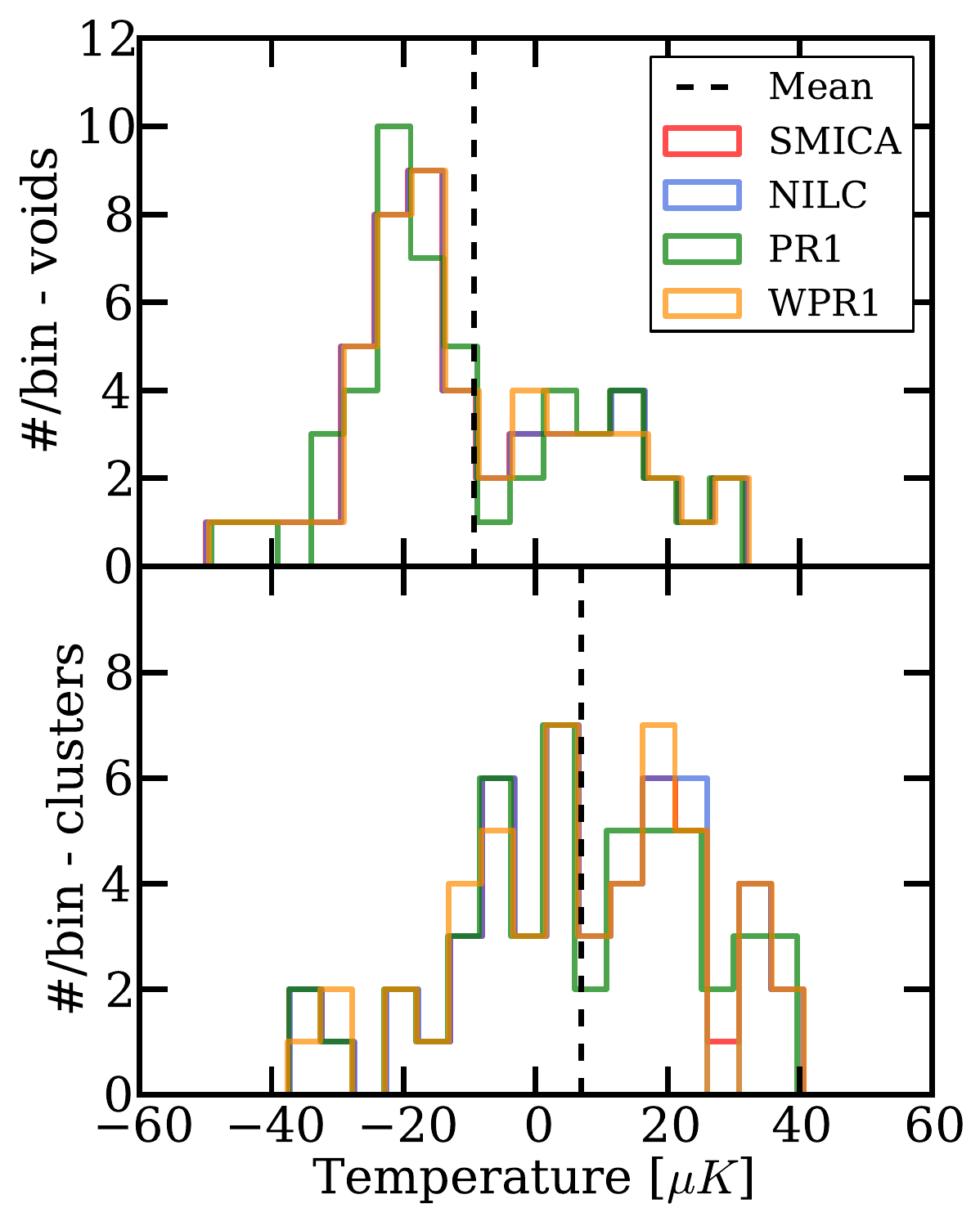}	
		\caption{Histograms of pixel temperatures centered on superstructures identified by GNS08, measured using 4 different foreground-cleaned filtered CMB maps. Top panel: measured temperatures at locations of voids in the GNS08 catalog; the dashed vertical line indicates the mean temperature. Bottom panel: the same  
for locations of clusters.}
	\label{fig:steps}
\end{figure}

We process these four maps in the same fashion:
\begin{itemize}
\item we apply a gaussian beam smoothing in harmonic space to the map defined as $B_{\ell} = B_{\ell}(30^{\prime})/B_{\ell}(\rm map)$ 
where $B_{\ell}(\rm map)$ is the effective beam of the released map; this allows us to take into account for the finite resolution 
of the instrument, and hence matching the overall smoothing applied to the simulated maps. 
We also filter out the small-scale fluctuations by setting the harmonic coefficients of the map $a_{\ell m }=0$ for $\ell>800$; 
\item the preprocessed map is then masked using the released \verb|Planck| mask \verb|U73|, avoiding contaminations from bright point sources;
\item the masked map is filtered in harmonic space using the compensated top-hat filter $F_{\ell}$ and repixelized to \verb|NSIDE|=$256$; 
\item we read the temperature values of the pixels corresponding to the cluster/void positions used in GNS08 \footnote{\url{http://ifa.hawaii.edu/cosmowave/supervoids/publications-and-data/}}. 
\end{itemize}

\begin{figure*}[t!]
	\centering
		\includegraphics[scale=0.5]{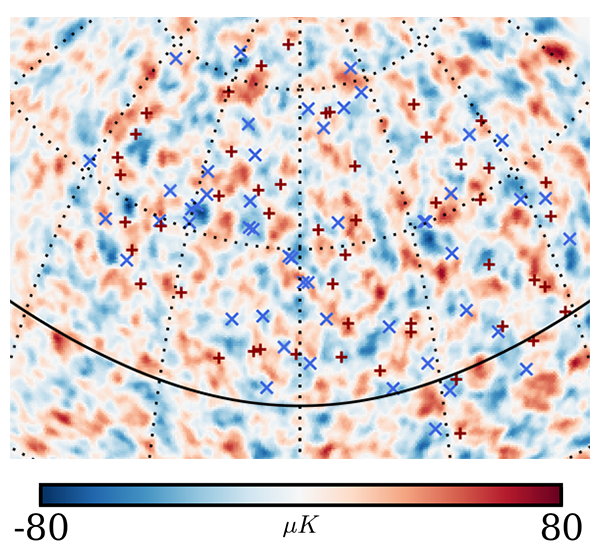}
		\caption{The filtered SMICA-Planck CMB temperature map, in a Mollweide projection in ecliptic coordinates. The galactic region and point sources have been masked with the U73-Planck mask. The resolution of the HEALPIX maps is NSIDE$=256$. The locations of superclusters (red ``+'') and supervoids (blue ``x'') from the GNS08 catalog are also shown.}
	\label{fig:map}
\end{figure*}

Fig.~\ref{fig:map} shows the filtered \verb|SMICA| map in a Mollweide projection in ecliptic coordinates; superstructure locations from GNS08 are marked. In Fig.~\ref{fig:steps}, we plot the histogram of the temperature values for voids and clusters separately for the four analyzed maps. The measured values are used to calculate the quantities $\overline{T_c}$, $\overline{T_h}$ and $\overline{T_m}$ given in Table~\ref{table:maps_res}. 
\begin{table}[t!] 
	\caption{Mean temperature deviations for GNS08 cluster and void locations, for four temperature maps with different foreground cleaning procedures.
We estimate the mean and standard deviation $\sigma_{FG}$ from the four different maps.}
\centering  
\begin{tabular}{c | c | c | c}
\hline\hline
&&&\\
Map & $\quad$ $\overline{T_h}$ $[\mu K]$ $\quad$ & $\quad$ $\overline{T_c}$ $[\mu K]$ $\quad$ & $\quad$ $\overline{T_m}$ $[\mu K]$\\ 
\hline\hline 
NILC & $ 6.9 $ & $-9.4$ & $8.1$\\
SMICA & $ 7.0 $ & $-9.4$ & $8.2$\\
PR1 & $ 6.9 $ & $-9.3$ & $8.1$\\
WPR1 & $ 6.9 $ & $-9.2$ & $8.0$\\
\hline\hline
MEAN & $6.89$ & $-9.33$ & $8.11$\\
$\sigma_{FG}$& $0.01$ & $0.09$ & $0.04$\\
\hline\hline
\end{tabular} 
\label{table:maps_res} 
\end{table} 
Different component separation methods quantify the effects of residual foreground contamination. We measure the fluctuations of the average temperature signal for different maps and use the variance of these fluctuations $\sigma_{\rm FG}$ as an estimate of the error due to foregrounds. The temperature values are extremely stable and fluctuations are always within 
$1\%$ (see also Fig.~\ref{fig:steps}), suggesting that the temperature variations are predominantly cosmological. Our mean peak temperature values are smaller than those
reported by GNS08 and PLK13 by around 1.5 $\mu$K, which is within the 1$\sigma$ uncertainty. Such a difference is driven mainly by details of the filtering procedure. 

The results of our simulations and our measured signals, shown in Fig.~\ref{fig:scatter_plot} and Fig.~\ref{fig:hist_plot}, can be summarized as
\begin{itemize}
\item The departure of the measurements from a null signal has decreased somewhat compared to previous analyses. It corresponds to a detection significance of $2.2\sigma$, $3.0\sigma$ and $3.5\sigma$ for clusters, voids and combined, respectively;
\item The measurements are higher than the expected maximum signal in $\Lambda$CDM cosmology at a level of $1.5\sigma$, $2.3\sigma$ and $2.5\sigma$ for clusters, voids and combined, respectively;
\newpage
\item The asymmetry between the measured signal for voids and clusters is not statistically significant, being smaller than $1\sigma$.
\end{itemize}
For these estimates, we consider foregrounds contamination and cosmic variance from simulations to be uncorrelated; hence we take $\sigma_{\rm tot} = \sqrt{\sigma^2_{\rm FG}+\sigma^2_{\rm sim}}$, but the residual foreground error is small compared to the cosmic variance uncertainty.

\section{Discussion\label{secV}}
 
Our analysis confirms both the size of the stacked late-ISW signal seen by GNS08 and PLK13, 
and theoretical predictions for $\Lambda$CDM models by FHN13 and HMS13. 
By using several maps with different foreground subtraction
methods, we demonstrate that foreground residuals contribute negligible uncertainty to the measured
signal. The theoretical modeling, using correlated Gaussian random fields, is far simpler than
previous analyses using N-body simulations, showing that the predicted signal has no
significant systematic error arising from insufficient box size or other subtleties of the simulations. 
Our calculations also include the correlations between the late-ISW signal
and other sources of microwave temperature anisotropies, which mildly increases the theoretical mean signal
while also increasing the statistical uncertainty.
We find a stacked late-ISW signal which is different from null at $3.5\sigma$ significance, 
and a discrepancy between the predicted and observed signal of $2.5\sigma$ in
Planck sky maps at the peak and void locations determined by 
GNS08 from SDSS data in the redshift range $0.4<z<0.75$.

The statistic used in this work is the mean value at the sky locations of the 50 highest positive
and lowest negative peaks in the late-ISW signal, assumed to be traced by structures and voids in a large-scale structure
survey. In simulations, the late-ISW peaks can be identified directly, and the 50 highest peaks in a given
sky region are known precisely. When analyzing large-scale structure data, peak identification will
not be perfectly efficient: some of the actual 50 largest extrema in the late-ISW signal may be missed
in favor of others which have lower amplitude. Thus the observed signal will necessarily be
biased low. The observed discrepancy between observation and theory has the observed
signal high compared to the prediction, so any systematic error in cluster identification has
reduced this discrepancy. In other words, our observed discrepancy is a lower limit to the
actual discrepancy, which may be larger than $2.5\sigma$ due to the identified clusters and voids
being imperfect tracers of the late-ISW temperature distribution. In reality, the total late-ISW signal is the
superposition of signals from very large numbers of voids and clusters, and it is not clear the extent to which
the largest voids and clusters individually produce local peaks in the filtered late-ISW map.
Since our predicted maximum signal is consistent with that from N-body simulations, 
it seems likely that large structures do actually produce local peaks in the filtered late-ISW map.
In the limit that the void and cluster locations from GNS08 do not
correlate at all with peaks in the late-ISW distribution, the model signal will be zero; but
then the mean signal at the GNS08 locations is $3.5\sigma$ away from the expected null signal.
 
The uncertainty in the difference between the observed signal and the theoretical maximum signal
is dominated by the primary temperature anisotropies which are uncorrelated with the late-ISW signal. 
When stacking at late-ISW peak locations, these primary fluctuations average to zero, with a Poisson error.
This uncertainty can be reduced only by including more peak locations in the average. The current
analysis uses late-ISW tracers from around 20\% of the sky, in a specific redshift range. Using the same
analysis with a half-sky survey at the same cluster and void threshold level 
will increase the number of voids and cluster locations by a factor of 2, reducing the Poisson error
by a factor of $\sqrt{2}$ and potentially increasing the detection significance of an underlying
signal discrepancy from $2.5\sigma$ to $3.5\sigma$. Extending the redshift range to lower $z$,
where the late-ISW effect is stronger for a given structure in standard $\Lambda$CDM models, can further increase the census of
clusters and voids, potentially pushing the discrepancy to greater than $4\sigma$. However, complications
at lower redshifts arise due to differing angular sizes of voids on the sky. A stacking analysis
at locations of lower-redshift SDSS voids has seen no signal clearly different from null \cite{ilic2013}, 
suggesting that the discrepancy here and in GNS08 may be due to noise. Upcoming optical
surveys like Skymapper \cite{skymapper} and LSST \cite{lsst} promise a substantial expansion in the 
census of voids and clusters suitable for late-ISW peak analysis. 
 
If the discrepancy is confirmed with increased statistical significance by future data, 
this would suggest that the late-ISW peak signal
is larger than in the standard $\Lambda$CDM model. Since the clusters and voids considered are on
very large scales, they are in the linear perturbation regime, and the physics determining
their late-ISW signal is simple, so it is unlikely that the theoretical signal in $\Lambda$CDM is being
computed incorrectly. While the association of voids or clusters with peaks in the late-ISW distribution
is challenging, any inefficiency in this process will only increase the discrepancy between theory
and measurement. The remaining possibility would be that the assumed expansion history in $\Lambda$CDM
is incorrect, and that the discrepancy indicates expansion dynamics different from that in models
with a cosmological constant. Any such modification must change the peak
statistics of the late-ISW temperature component while remaining within the bounds on the
total temperature power spectrum at large scales, and must be consistent with measurements
of the cross correlation between galaxies and microwave temperature. Given the
limited number of observational handles on the dark energy phenomenon, further work to understand
the mean peak late-ISW signal in current data, and its measurement with future larger galaxy surveys,
is of pressing interest.\\ 

\begin{acknowledgments}
The authors thank Ben Granett,  Seshadri Nadathur, Mark Neyrinck, and Uros Seljak for useful discussions. SA is grateful to Federico Bianchini 
for introducing him to the CLASS code and Julien Lesgourgues for helpful suggestions on modifying it. SA and AK are supported in part by 
the National Science Foundation under grant NSF-AST-1312380. AK thanks the International Center for Theoretical Physics, where this work was initiated during a stimulating conference.
We have used the \verb|HEALPix| package \cite{HEALPix} for map pixelization and the NASA Astrophysical Data System for bibliographic information.
\end{acknowledgments}

\bibliography{ISW_Stack_AKW_Oct_2014.bib}

\begin{thebibliography}{10}%
\makeatletter
\providecommand \@ifxundefined [1]{%
 \ifx #1\undefined \expandafter \@firstoftwo
 \else \expandafter \@secondoftwo
\fi
}%
\providecommand \@ifnum [1]{%
 \ifnum #1\expandafter \@firstoftwo
 \else \expandafter \@secondoftwo
\fi
}%
\providecommand \enquote [1]{``#1''}%
\providecommand \bibnamefont  [1]{#1}%
\providecommand \bibfnamefont [1]{#1}%
\providecommand \citenamefont [1]{#1}%
\providecommand\href[0]{\@sanitize\@href}%
\providecommand\@href[1]{\endgroup\@@startlink{#1}\endgroup\@@href}%
\providecommand\@@href[1]{#1\@@endlink}%
\providecommand \@sanitize [0]{\begingroup\catcode`\&12\catcode`\#12\relax}%
\@ifxundefined \pdfoutput {\@firstoftwo}{%
 \@ifnum{\z@=\pdfoutput}{\@firstoftwo}{\@secondoftwo}%
}{%
 \providecommand\@@startlink[1]{\leavevmode\special{html:<a href="#1">}}%
 \providecommand\@@endlink[0]{\special{html:</a>}}%
}{%
 \providecommand\@@startlink[1]{%
  \leavevmode
  \pdfstartlink
   attr{/Border[0 0 1 ]/H/I/C[0 1 1]}%
   user{/Subtype/Link/A<</Type/Action/S/URI/URI(#1)>>}%
  \relax
 }%
 \providecommand\@@endlink[0]{\pdfendlink}%
}%
\providecommand \url  [0]{\begingroup\@sanitize \@url }%
\providecommand \@url [1]{\endgroup\@href {#1}{\urlprefix}}%
\providecommand \urlprefix [0]{URL }%
\providecommand \Eprint[0]{\href }%
\@ifxundefined \urlstyle {%
  \providecommand \doi [1]{doi:\discretionary{}{}{}#1}%
}{%
  \providecommand \doi [0]{doi:\discretionary{}{}{}\begingroup
  \urlstyle{rm}\Url }%
}%
\providecommand \doibase [0]{http://dx.doi.org/}%
\providecommand \Doi[1]{\href{\doibase#1}}%
\providecommand \bibAnnote [3]{%
  \BibitemShut{#1}%
  \begin{quotation}\noindent
    \textsc{Key:}\ #2\\\textsc{Annotation:}\ #3%
  \end{quotation}%
}%
\providecommand \bibAnnoteFile [2]{%
  \IfFileExists{#2}{\bibAnnote {#1} {#2} {\input{#2}}}{}%
}%
\providecommand \typeout [0]{\immediate \write \m@ne }%
\providecommand \selectlanguage [0]{\@gobble}%
\providecommand \bibinfo [0]{\@secondoftwo}%
\providecommand \bibfield [0]{\@secondoftwo}%
\providecommand \translation [1]{[#1]}%
\providecommand \BibitemOpen[0]{}%
\providecommand \bibitemStop [0]{}%
\providecommand \bibitemNoStop [0]{.\EOS\space}%
\providecommand \EOS [0]{\spacefactor3000\relax}%
\providecommand \BibitemShut [1]{\csname bibitem#1\endcsname}%
\bibitem{Riess}%
  \BibitemOpen
  \bibfield{author}{%
  \bibinfo {author} {\bibfnamefont{A.~G.}\ \bibnamefont{{Riess}}}
  \emph{et~al.},\ }%
  \bibfield{journal}{%
  \Doi{10.1086/300499}{\bibinfo {journal} {Astronom.~J.}}\ }%
  \textbf{\bibinfo {volume} {116}},\ \bibinfo {pages} {1009} (\bibinfo {year}
  {1998}),\ \Eprint{http://arxiv.org/abs/astro-ph/9805201}{astro-ph/9805201}%
  \bibAnnoteFile{NoStop}{Riess}%
\bibitem{Perlmutter}%
  \BibitemOpen
  \bibfield{author}{%
  \bibinfo {author} {\bibfnamefont{S.}~\bibnamefont{{Perlmutter}}}
  \emph{et~al.},\ }%
  \bibfield{journal}{%
  \Doi{10.1086/307221}{\bibinfo {journal} {\apj}}\ }%
  \textbf{\bibinfo {volume} {517}},\ \bibinfo {pages} {565} (\bibinfo {year}
  {1999}),\ \Eprint{http://arxiv.org/abs/astro-ph/9812133}{astro-ph/9812133}%
  \bibAnnoteFile{NoStop}{Perlmutter}%
\bibitem{Sherwin}%
  \BibitemOpen
  \bibfield{author}{%
  \bibinfo {author} {\bibfnamefont{B.~D.}\ \bibnamefont{{Sherwin}}}, \bibinfo
  {author} {\bibfnamefont{J.}~\bibnamefont{{Dunkley}}}, \bibinfo {author}
  {\bibfnamefont{S.}~\bibnamefont{{Das}}}, \emph{et~al.},\ }%
  \bibfield{journal}{%
  \Doi{10.1103/PhysRevLett.107.021302}{\bibinfo {journal} {\prl}}\ }%
  \textbf{\bibinfo {volume} {107}},\ \bibinfo {eid} {021302} (\bibinfo {year}
  {2011}),\ \Eprint{http://arxiv.org/abs/1105.0419}{arXiv:1105.0419
  [astro-ph.CO]}%
  \bibAnnoteFile{NoStop}{Sherwin}%
\bibitem{VanEngelen}%
  \BibitemOpen
  \bibfield{author}{%
  \bibinfo {author} {\bibfnamefont{A.}~\bibnamefont{{van Engelen}}}, \bibinfo
  {author} {\bibfnamefont{R.}~\bibnamefont{{Keisler}}}, \bibinfo {author}
  {\bibfnamefont{O.}~\bibnamefont{{Zahn}}}, \emph{et~al.},\ }%
  \bibfield{journal}{%
  \Doi{10.1088/0004-637X/756/2/142}{\bibinfo {journal} {\apj}}\ }%
  \textbf{\bibinfo {volume} {756}},\ \bibinfo {eid} {142} (\bibinfo {year}
  {2012}),\ \Eprint{http://arxiv.org/abs/1202.0546}{arXiv:1202.0546
  [astro-ph.CO]}%
  \bibAnnoteFile{NoStop}{VanEngelen}%
\bibitem{PLANCK_PARAMS}%
  \BibitemOpen
  \bibfield{author}{%
  \bibinfo {author} {\bibnamefont{{Planck Collaboration}}}}%
   (\bibinfo {year} {2013}),\
  \Eprint{http://arxiv.org/abs/1303.5076}{arXiv:1303.5076 [astro-ph.CO]}%
  \bibAnnoteFile{NoStop}{PLANCK_PARAMS}%
\bibitem{BAO}%
  \BibitemOpen
  \bibfield{author}{%
  \bibinfo {author} {\bibfnamefont{T.}~\bibnamefont{{Delubac}}} \emph{et~al.}}%
   (\bibinfo {year} {2014}),\
  \Eprint{http://arxiv.org/abs/1404.1801}{arXiv:1404.1801}%
  \bibAnnoteFile{NoStop}{BAO}%
\bibitem{SachsWolfe}%
  \BibitemOpen
  \bibfield{author}{%
  \bibinfo {author} {\bibfnamefont{R.~K.}\ \bibnamefont{{Sachs}}}\ and\
  \bibinfo {author} {\bibfnamefont{A.~M.}\ \bibnamefont{{Wolfe}}},\ }%
  \bibfield{journal}{%
  \Doi{10.1086/148982}{\bibinfo {journal} {\apj}}\ }%
  \textbf{\bibinfo {volume} {147}},\ \bibinfo {pages} {73} (\bibinfo {year}
  {1967})%
  \bibAnnoteFile{NoStop}{SachsWolfe}%
\bibitem{Kamionkowski}%
  \BibitemOpen
  \bibfield{author}{%
  \bibinfo {author} {\bibfnamefont{M.}~\bibnamefont{{Kamionkowski}}}\ and\
  \bibinfo {author} {\bibfnamefont{D.~N.}\ \bibnamefont{{Spergel}}},\ }%
  \bibfield{journal}{%
  \Doi{10.1086/174543}{\bibinfo {journal} {\apj}}\ }%
  \textbf{\bibinfo {volume} {432}},\ \bibinfo {pages} {7} (\bibinfo {year}
  {1994}),\ \Eprint{http://arxiv.org/abs/astro-ph/9312017}{astro-ph/9312017}%
  \bibAnnoteFile{NoStop}{Kamionkowski}%
\bibitem{Giannantonio08}%
  \BibitemOpen
  \bibfield{author}{%
  \bibinfo {author} {\bibfnamefont{T.}~\bibnamefont{{Giannantonio}}}, \bibinfo
  {author} {\bibfnamefont{R.}~\bibnamefont{{Scranton}}}, \bibinfo {author}
  {\bibfnamefont{R.~G.}\ \bibnamefont{{Crittenden}}}, \bibinfo {author}
  {\bibfnamefont{R.~C.}\ \bibnamefont{{Nichol}}}, \bibinfo {author}
  {\bibfnamefont{S.~P.}\ \bibnamefont{{Boughn}}}, \bibinfo {author}
  {\bibfnamefont{A.~D.}\ \bibnamefont{{Myers}}},\ and\ \bibinfo {author}
  {\bibfnamefont{G.~T.}\ \bibnamefont{{Richards}}},\ }%
  \bibfield{journal}{%
  \Doi{10.1103/PhysRevD.77.123520}{\bibinfo {journal} {\prd}}\ }%
  \textbf{\bibinfo {volume} {77}},\ \bibinfo {eid} {123520} (\bibinfo {year}
  {2008}),\ \Eprint{http://arxiv.org/abs/0801.4380}{arXiv:0801.4380}%
  \bibAnnoteFile{NoStop}{Giannantonio08}%
\bibitem{Crittenden}%
  \BibitemOpen
  \bibfield{author}{%
  \bibinfo {author} {\bibfnamefont{R.~G.}\ \bibnamefont{{Crittenden}}}\ and\
  \bibinfo {author} {\bibfnamefont{N.}~\bibnamefont{{Turok}}},\ }%
  \bibfield{journal}{%
  \Doi{10.1103/PhysRevLett.76.575}{\bibinfo {journal} {\prl}}\ }%
  \textbf{\bibinfo {volume} {76}},\ \bibinfo {pages} {575} (\bibinfo {year}
  {1996}),\ \Eprint{http://arxiv.org/abs/astro-ph/9510072}{astro-ph/9510072}%
  \bibAnnoteFile{NoStop}{Crittenden}%
\bibitem{Fosalba}%
  \BibitemOpen
  \bibfield{author}{%
  \bibinfo {author} {\bibfnamefont{P.}~\bibnamefont{{Fosalba}}}\ and\ \bibinfo
  {author} {\bibfnamefont{E.}~\bibnamefont{{Gazta{\~n}aga}}},\ }%
  \bibfield{journal}{%
  \Doi{10.1111/j.1365-2966.2004.07837.x}{\bibinfo {journal} {MNRAS}}\ }%
  \textbf{\bibinfo {volume} {350}},\ \bibinfo {pages} {L37} (\bibinfo {year}
  {2004}),\ \Eprint{http://arxiv.org/abs/astro-ph/0305468}{astro-ph/0305468}%
  \bibAnnoteFile{NoStop}{Fosalba}%
\bibitem{Francis}%
  \BibitemOpen
  \bibfield{author}{%
  \bibinfo {author} {\bibfnamefont{C.~L.}\ \bibnamefont{{Francis}}}\ and\
  \bibinfo {author} {\bibfnamefont{J.~A.}\ \bibnamefont{{Peacock}}},\ }%
  \bibfield{journal}{%
  \Doi{10.1111/j.1365-2966.2010.16278.x}{\bibinfo {journal} {MNRAS}}\ }%
  \textbf{\bibinfo {volume} {406}},\ \bibinfo {pages} {2} (\bibinfo {year}
  {2010}),\ \Eprint{http://arxiv.org/abs/0909.2494}{arXiv:0909.2494
  [astro-ph.CO]}%
  \bibAnnoteFile{NoStop}{Francis}%
\bibitem{Afshordi}%
  \BibitemOpen
  \bibfield{author}{%
  \bibinfo {author} {\bibfnamefont{N.}~\bibnamefont{{Afshordi}}}, \bibinfo
  {author} {\bibfnamefont{Y.-S.}\ \bibnamefont{{Loh}}},\ and\ \bibinfo {author}
  {\bibfnamefont{M.~A.}\ \bibnamefont{{Strauss}}},\ }%
  \bibfield{journal}{%
  \Doi{10.1103/PhysRevD.69.083524}{\bibinfo {journal} {\prd}}\ }%
  \textbf{\bibinfo {volume} {69}},\ \bibinfo {eid} {083524} (\bibinfo {year}
  {2004}),\ \Eprint{http://arxiv.org/abs/astro-ph/0308260}{astro-ph/0308260}%
  \bibAnnoteFile{NoStop}{Afshordi}%
\bibitem{Boughn}%
  \BibitemOpen
  \bibfield{author}{%
  \bibinfo {author} {\bibfnamefont{S.}~\bibnamefont{{Boughn}}}\ and\ \bibinfo
  {author} {\bibfnamefont{R.}~\bibnamefont{{Crittenden}}},\ }%
  \bibfield{journal}{%
  \Doi{10.1038/nature02139}{\bibinfo {journal} {Nature}}\ }%
  \textbf{\bibinfo {volume} {427}},\ \bibinfo {pages} {45} (\bibinfo {year}
  {2004}),\ \Eprint{http://arxiv.org/abs/astro-ph/0305001}{astro-ph/0305001}%
  \bibAnnoteFile{NoStop}{Boughn}%
\bibitem{Nolta}%
  \BibitemOpen
  \bibfield{author}{%
  \bibinfo {author} {\bibfnamefont{M.~R.}\ \bibnamefont{{Nolta}}}
  \emph{et~al.},\ }%
  \bibfield{journal}{%
  \Doi{10.1086/386536}{\bibinfo {journal} {\apj}}\ }%
  \textbf{\bibinfo {volume} {608}},\ \bibinfo {pages} {10} (\bibinfo {year}
  {2004}),\ \Eprint{http://arxiv.org/abs/astro-ph/0305097}{astro-ph/0305097}%
  \bibAnnoteFile{NoStop}{Nolta}%
\bibitem{Padmanabhan}%
  \BibitemOpen
  \bibfield{author}{%
  \bibinfo {author} {\bibfnamefont{N.}~\bibnamefont{{Padmanabhan}}}, \bibinfo
  {author} {\bibfnamefont{C.~M.}\ \bibnamefont{{Hirata}}}, \bibinfo {author}
  {\bibfnamefont{U.}~\bibnamefont{{Seljak}}}, \bibinfo {author}
  {\bibfnamefont{D.~J.}\ \bibnamefont{{Schlegel}}}, \bibinfo {author}
  {\bibfnamefont{J.}~\bibnamefont{{Brinkmann}}},\ and\ \bibinfo {author}
  {\bibfnamefont{D.~P.}\ \bibnamefont{{Schneider}}},\ }%
  \bibfield{journal}{%
  \Doi{10.1103/PhysRevD.72.043525}{\bibinfo {journal} {\prd}}\ }%
  \textbf{\bibinfo {volume} {72}},\ \bibinfo {eid} {043525} (\bibinfo {year}
  {2005}),\ \Eprint{http://arxiv.org/abs/astro-ph/0410360}{astro-ph/0410360}%
  \bibAnnoteFile{NoStop}{Padmanabhan}%
\bibitem{Rassat}%
  \BibitemOpen
  \bibfield{author}{%
  \bibinfo {author} {\bibfnamefont{A.}~\bibnamefont{{Rassat}}}, \bibinfo
  {author} {\bibfnamefont{K.}~\bibnamefont{{Land}}}, \bibinfo {author}
  {\bibfnamefont{O.}~\bibnamefont{{Lahav}}},\ and\ \bibinfo {author}
  {\bibfnamefont{F.~B.}\ \bibnamefont{{Abdalla}}},\ }%
  \bibfield{journal}{%
  \Doi{10.1111/j.1365-2966.2007.11538.x}{\bibinfo {journal} {MNRAS}}\ }%
  \textbf{\bibinfo {volume} {377}},\ \bibinfo {pages} {1085} (\bibinfo {year}
  {2007}),\ \Eprint{http://arxiv.org/abs/astro-ph/0610911}{astro-ph/0610911}%
  \bibAnnoteFile{NoStop}{Rassat}%
\bibitem{Sawangwit}%
  \BibitemOpen
  \bibfield{author}{%
  \bibinfo {author} {\bibfnamefont{U.}~\bibnamefont{{Sawangwit}}}, \bibinfo
  {author} {\bibfnamefont{T.}~\bibnamefont{{Shanks}}}, \bibinfo {author}
  {\bibfnamefont{R.~D.}\ \bibnamefont{{Cannon}}}, \bibinfo {author}
  {\bibfnamefont{S.~M.}\ \bibnamefont{{Croom}}}, \bibinfo {author}
  {\bibfnamefont{N.~P.}\ \bibnamefont{{Ross}}},\ and\ \bibinfo {author}
  {\bibfnamefont{D.~A.}\ \bibnamefont{{Wake}}},\ }%
  \bibfield{journal}{%
  \Doi{10.1111/j.1365-2966.2009.16054.x}{\bibinfo {journal} {MNRAS}}\ }%
  \textbf{\bibinfo {volume} {402}},\ \bibinfo {pages} {2228} (\bibinfo {year}
  {2010}),\ \Eprint{http://arxiv.org/abs/0911.1352}{arXiv:0911.1352
  [astro-ph.CO]}%
  \bibAnnoteFile{NoStop}{Sawangwit}%
\bibitem{Vielva}%
  \BibitemOpen
  \bibfield{author}{%
  \bibinfo {author} {\bibfnamefont{P.}~\bibnamefont{{Vielva}}}, \bibinfo
  {author} {\bibfnamefont{E.}~\bibnamefont{{Mart{\'{\i}}nez-Gonz{\'a}lez}}},\
  and\ \bibinfo {author} {\bibfnamefont{M.}~\bibnamefont{{Tucci}}},\ }%
  \bibfield{journal}{%
  \Doi{10.1111/j.1365-2966.2005.09764.x}{\bibinfo {journal} {MNRAS}}\ }%
  \textbf{\bibinfo {volume} {365}},\ \bibinfo {pages} {891} (\bibinfo {year}
  {2006}),\ \Eprint{http://arxiv.org/abs/astro-ph/0408252}{astro-ph/0408252}%
  \bibAnnoteFile{NoStop}{Vielva}%
\bibitem{Cabre}%
  \BibitemOpen
  \bibfield{author}{%
  \bibinfo {author} {\bibfnamefont{A.}~\bibnamefont{{Cabr{\'e}}}}, \bibinfo
  {author} {\bibfnamefont{E.}~\bibnamefont{{Gazta{\~n}aga}}}, \bibinfo {author}
  {\bibfnamefont{M.}~\bibnamefont{{Manera}}}, \bibinfo {author}
  {\bibfnamefont{P.}~\bibnamefont{{Fosalba}}},\ and\ \bibinfo {author}
  {\bibfnamefont{F.}~\bibnamefont{{Castander}}},\ }%
  \bibfield{journal}{%
  \Doi{10.1111/j.1745-3933.2006.00218.x}{\bibinfo {journal} {MNRAS}}\ }%
  \textbf{\bibinfo {volume} {372}},\ \bibinfo {pages} {L23} (\bibinfo {year}
  {2006}),\ \Eprint{http://arxiv.org/abs/astro-ph/0603690}{astro-ph/0603690}%
  \bibAnnoteFile{NoStop}{Cabre}%
\bibitem{Ho}%
  \BibitemOpen
  \bibfield{author}{%
  \bibinfo {author} {\bibfnamefont{S.}~\bibnamefont{{Ho}}}, \bibinfo {author}
  {\bibfnamefont{C.}~\bibnamefont{{Hirata}}}, \bibinfo {author}
  {\bibfnamefont{N.}~\bibnamefont{{Padmanabhan}}}, \bibinfo {author}
  {\bibfnamefont{U.}~\bibnamefont{{Seljak}}},\ and\ \bibinfo {author}
  {\bibfnamefont{N.}~\bibnamefont{{Bahcall}}},\ }%
  \bibfield{journal}{%
  \Doi{10.1103/PhysRevD.78.043519}{\bibinfo {journal} {\prd}}\ }%
  \textbf{\bibinfo {volume} {78}},\ \bibinfo {eid} {043519} (\bibinfo {year}
  {2008}),\ \Eprint{http://arxiv.org/abs/0801.0642}{arXiv:0801.0642}%
  \bibAnnoteFile{NoStop}{Ho}%
\bibitem{Planck_ISW}%
  \BibitemOpen
  \bibfield{author}{%
  \bibinfo {author} {\bibnamefont{{Planck Collaboration}}}}%
   (\bibinfo {year} {2013}),\
  \Eprint{http://arxiv.org/abs/1303.5079}{arXiv:1303.5079 [astro-ph.CO]}%
  \bibAnnoteFile{NoStop}{Planck_ISW}%
\bibitem{Monteagudo13}%
  \BibitemOpen
  \bibfield{author}{%
  \bibinfo {author}
  {\bibfnamefont{C.}~\bibnamefont{{Hern{\'a}ndez-Monteagudo}}}, \bibinfo
  {author} {\bibfnamefont{A.~J.}\ \bibnamefont{{Ross}}}, \emph{et~al.},\ }%
  \bibfield{journal}{%
  \Doi{10.1093/mnras/stt2312}{\bibinfo {journal} {MNRAS}}\ }%
  \textbf{\bibinfo {volume} {438}},\ \bibinfo {pages} {1724} (\bibinfo {year}
  {2014}),\ \Eprint{http://arxiv.org/abs/1303.4302}{arXiv:1303.4302
  [astro-ph.CO]}%
  \bibAnnoteFile{NoStop}{Monteagudo13}%
\bibitem{Kovacs}%
  \BibitemOpen
  \bibfield{author}{%
  \bibinfo {author} {\bibfnamefont{A.}~\bibnamefont{{Kov{\'a}cs}}}, \bibinfo
  {author} {\bibfnamefont{I.}~\bibnamefont{{Szapudi}}}, \bibinfo {author}
  {\bibfnamefont{B.~R.}\ \bibnamefont{{Granett}}},\ and\ \bibinfo {author}
  {\bibfnamefont{Z.}~\bibnamefont{{Frei}}},\ }%
  \bibfield{journal}{%
  \Doi{10.1093/mnrasl/slt002}{\bibinfo {journal} {MNRAS}}\ }%
  \textbf{\bibinfo {volume} {431}},\ \bibinfo {pages} {L28} (\bibinfo {year}
  {2013}),\ \Eprint{http://arxiv.org/abs/1301.0475}{arXiv:1301.0475
  [astro-ph.CO]}%
  \bibAnnoteFile{NoStop}{Kovacs}%
\bibitem{Ferraro}%
  \BibitemOpen
  \bibfield{author}{%
  \bibinfo {author} {\bibfnamefont{S.}~\bibnamefont{{Ferraro}}}, \bibinfo
  {author} {\bibfnamefont{B.~D.}\ \bibnamefont{{Sherwin}}},\ and\ \bibinfo
  {author} {\bibfnamefont{D.~N.}\ \bibnamefont{{Spergel}}}}%
   (\bibinfo {year} {2014}),\
  \Eprint{http://arxiv.org/abs/1401.1193}{arXiv:1401.1193 [astro-ph.CO]}%
  \bibAnnoteFile{NoStop}{Ferraro}%
\bibitem{Nishizawa}%
  \BibitemOpen
  \bibfield{author}{%
  \bibinfo {author} {\bibfnamefont{A.~J.}\ \bibnamefont{{Nishizawa}}}}%
   (\bibinfo {year} {2014}),\
  \Eprint{http://arxiv.org/abs/1404.5102}{arXiv:1404.5102}%
  \bibAnnoteFile{NoStop}{Nishizawa}%
\bibitem{Granett}%
  \BibitemOpen
  \bibfield{author}{%
  \bibinfo {author} {\bibfnamefont{B.~R.}\ \bibnamefont{{Granett}}}, \bibinfo
  {author} {\bibfnamefont{M.~C.}\ \bibnamefont{{Neyrinck}}},\ and\ \bibinfo
  {author} {\bibfnamefont{I.}~\bibnamefont{{Szapudi}}},\ }%
  \bibfield{journal}{%
  \Doi{10.1086/591670}{\bibinfo {journal} {\apj}}\ }%
  \textbf{\bibinfo {volume} {683}},\ \bibinfo {pages} {L99} (\bibinfo {year}
  {2008}),\ \Eprint{http://arxiv.org/abs/0805.3695}{arXiv:0805.3695}%
  \bibAnnoteFile{NoStop}{Granett}%
\bibitem{hinshaw09}%
  \BibitemOpen
  \bibfield{author}{%
  \bibinfo {author} {\bibfnamefont{G.}~\bibnamefont{{Hinshaw}}}, \bibinfo
  {author} {\bibfnamefont{J.~L.}\ \bibnamefont{{Weiland}}}, \bibinfo {author}
  {\bibfnamefont{R.~S.}\ \bibnamefont{{Hill}}}, \bibinfo {author}
  {\bibfnamefont{N.}~\bibnamefont{{Odegard}}}, \bibinfo {author}
  {\bibfnamefont{D.}~\bibnamefont{{Larson}}}, \emph{et~al.},\ }%
  \bibfield{journal}{%
  \Doi{10.1088/0067-0049/180/2/225}{\bibinfo {journal} {ApJSÊ}}\ }%
  \textbf{\bibinfo {volume} {180}},\ \bibinfo {pages} {225} (\bibinfo {year}
  {2009}),\ \Eprint{http://arxiv.org/abs/0803.0732}{arXiv:0803.0732}%
  \bibAnnoteFile{NoStop}{hinshaw09}%
\bibitem{SDSS6}%
  \BibitemOpen
  \bibfield{author}{%
  \bibinfo {author} {\bibfnamefont{J.~K.}\ \bibnamefont{{Adelman-McCarthy}}}
  \emph{et~al.},\ }%
  \bibfield{journal}{%
  \Doi{10.1086/524984}{\bibinfo {journal} {ApJSÊ}}\ }%
  \textbf{\bibinfo {volume} {175}},\ \bibinfo {pages} {297} (\bibinfo {year}
  {2008}),\ \Eprint{http://arxiv.org/abs/0707.3413}{arXiv:0707.3413}%
  \bibAnnoteFile{NoStop}{SDSS6}%
\bibitem{Monteagudo}%
  \BibitemOpen
  \bibfield{author}{%
  \bibinfo {author}
  {\bibfnamefont{C.}~\bibnamefont{{Hern{\'a}ndez-Monteagudo}}}\ and\ \bibinfo
  {author} {\bibfnamefont{R.~E.}\ \bibnamefont{{Smith}}},\ }%
  \bibfield{journal}{%
  \Doi{10.1093/mnras/stt1322}{\bibinfo {journal} {MNRAS}}\ }%
  \textbf{\bibinfo {volume} {435}},\ \bibinfo {pages} {1094} (\bibinfo {year}
  {2013}),\ \Eprint{http://arxiv.org/abs/1212.1174}{arXiv:1212.1174
  [astro-ph.CO]}%
  \bibAnnoteFile{NoStop}{Monteagudo}%
\bibitem{GNS08Sup}%
  \BibitemOpen
  \bibfield{author}{%
  \bibinfo {author} {\bibfnamefont{B.~R.}\ \bibnamefont{{Granett}}}, \bibinfo
  {author} {\bibfnamefont{M.~C.}\ \bibnamefont{{Neyrinck}}},\ and\ \bibinfo
  {author} {\bibfnamefont{I.}~\bibnamefont{{Szapudi}}}}%
   (\bibinfo {year} {2008}),\
  \Eprint{http://arxiv.org/abs/0805.2974}{arXiv:0805.2974}%
  \bibAnnoteFile{NoStop}{GNS08Sup}%
\bibitem{Nadathur}%
  \BibitemOpen
  \bibfield{author}{%
  \bibinfo {author} {\bibfnamefont{S.}~\bibnamefont{{Nadathur}}}, \bibinfo
  {author} {\bibfnamefont{S.}~\bibnamefont{{Hotchkiss}}},\ and\ \bibinfo
  {author} {\bibfnamefont{S.}~\bibnamefont{{Sarkar}}},\ }%
  \bibfield{journal}{%
  \Doi{10.1088/1475-7516/2012/06/042}{\bibinfo {journal} {JCAP}}\ }%
  \textbf{\bibinfo {volume} {6}},\ \bibinfo {eid} {042} (\bibinfo {year}
  {2012}),\ \Eprint{http://arxiv.org/abs/1109.4126}{arXiv:1109.4126
  [astro-ph.CO]}%
  \bibAnnoteFile{NoStop}{Nadathur}%
\bibitem{Flender}%
  \BibitemOpen
  \bibfield{author}{%
  \bibinfo {author} {\bibfnamefont{S.}~\bibnamefont{{Flender}}}, \bibinfo
  {author} {\bibfnamefont{S.}~\bibnamefont{{Hotchkiss}}},\ and\ \bibinfo
  {author} {\bibfnamefont{S.}~\bibnamefont{{Nadathur}}},\ }%
  \bibfield{journal}{%
  \Doi{10.1088/1475-7516/2013/02/013}{\bibinfo {journal} {JCAP}}\ }%
  \textbf{\bibinfo {volume} {2}},\ \bibinfo {eid} {013} (\bibinfo {year}
  {2013}),\ \Eprint{http://arxiv.org/abs/1212.0776}{arXiv:1212.0776
  [astro-ph.CO]}%
  \bibAnnoteFile{NoStop}{Flender}%
\bibitem{ReesSciama}%
  \BibitemOpen
  \bibfield{author}{%
  \bibinfo {author} {\bibfnamefont{M.~J.}\ \bibnamefont{{Rees}}}\ and\ \bibinfo
  {author} {\bibfnamefont{D.~W.}\ \bibnamefont{{Sciama}}},\ }%
  \bibfield{journal}{%
  \Doi{10.1038/217511a0}{\bibinfo {journal} {Nature}}\ }%
  \textbf{\bibinfo {volume} {217}},\ \bibinfo {pages} {511} (\bibinfo {year}
  {1968})%
  \bibAnnoteFile{NoStop}{ReesSciama}%
\bibitem{Cai}%
  \BibitemOpen
  \bibfield{author}{%
  \bibinfo {author} {\bibfnamefont{Y.-C.}\ \bibnamefont{{Cai}}}, \bibinfo
  {author} {\bibfnamefont{B.}~\bibnamefont{{Li}}}, \bibinfo {author}
  {\bibfnamefont{S.}~\bibnamefont{{Cole}}}, \bibinfo {author}
  {\bibfnamefont{C.~S.}\ \bibnamefont{{Frenk}}},\ and\ \bibinfo {author}
  {\bibfnamefont{M.}~\bibnamefont{{Neyrinck}}},\ }%
  \bibfield{journal}{%
  \Doi{10.1093/mnras/stu154}{\bibinfo {journal} {MNRAS}}\ }%
  \textbf{\bibinfo {volume} {439}},\ \bibinfo {pages} {2978} (\bibinfo {year}
  {2014}),\ \Eprint{http://arxiv.org/abs/1310.6986}{arXiv:1310.6986
  [astro-ph.CO]}%
  \bibAnnoteFile{NoStop}{Cai}%
\bibitem{Chen}%
  \BibitemOpen
  \bibfield{author}{%
  \bibinfo {author} {\bibfnamefont{B.}~\bibnamefont{{Chen}}}, \bibinfo {author}
  {\bibfnamefont{R.}~\bibnamefont{{Kantowski}}},\ and\ \bibinfo {author}
  {\bibfnamefont{X.}~\bibnamefont{{Dai}}},\ }%
  \bibfield{journal}{%
  \bibinfo {journal} {ArXiv e-prints}}%
   (\bibinfo {year} {2013}),\
  \Eprint{http://arxiv.org/abs/1310.6351}{arXiv:1310.6351}%
  \bibAnnoteFile{NoStop}{Chen}%
\bibitem{Papai10}%
  \BibitemOpen
  \bibfield{author}{%
  \bibinfo {author} {\bibfnamefont{P.}~\bibnamefont{{P{\'a}pai}}}, \bibinfo
  {author} {\bibfnamefont{I.}~\bibnamefont{{Szapudi}}},\ and\ \bibinfo {author}
  {\bibfnamefont{B.~R.}\ \bibnamefont{{Granett}}},\ }%
  \bibfield{journal}{%
  \Doi{10.1088/0004-637X/732/1/27}{\bibinfo {journal} {\apj}}\ }%
  \textbf{\bibinfo {volume} {732}},\ \bibinfo {eid} {27} (\bibinfo {year}
  {2011}),\ \Eprint{http://arxiv.org/abs/1012.3750}{arXiv:1012.3750
  [astro-ph.CO]}%
  \bibAnnoteFile{NoStop}{Papai10}%
\bibitem{Watson14}%
  \BibitemOpen
  \bibfield{author}{%
  \bibinfo {author} {\bibfnamefont{W.~A.}\ \bibnamefont{{Watson}}}
  \emph{et~al.},\ }%
  \bibfield{journal}{%
  \Doi{10.1093/mnras/stt2208}{\bibinfo {journal} {MNRAS}}\ }%
  \textbf{\bibinfo {volume} {438}},\ \bibinfo {pages} {412} (\bibinfo {year}
  {2014}),\ \Eprint{http://arxiv.org/abs/1307.1712}{arXiv:1307.1712}%
  \bibAnnoteFile{NoStop}{Watson14}%
\bibitem{Hotchkiss}%
  \BibitemOpen
  \bibfield{author}{%
  \bibinfo {author} {\bibfnamefont{S.}~\bibnamefont{{Hotchkiss}}}, \bibinfo
  {author} {\bibfnamefont{S.}~\bibnamefont{{Nadathur}}}, \bibinfo {author}
  {\bibfnamefont{S.}~\bibnamefont{{Gottl{\"o}ber}}}, \bibinfo {author}
  {\bibfnamefont{I.~T.}\ \bibnamefont{{Iliev}}}, \bibinfo {author}
  {\bibfnamefont{A.}~\bibnamefont{{Knebe}}}, \bibinfo {author}
  {\bibfnamefont{W.~A.}\ \bibnamefont{{Watson}}},\ and\ \bibinfo {author}
  {\bibfnamefont{G.}~\bibnamefont{{Yepes}}}}%
   (\bibinfo {year} {2014}),\
  \Eprint{http://arxiv.org/abs/1405.3552}{arXiv:1405.3552}%
  \bibAnnoteFile{NoStop}{Hotchkiss}%
\bibitem{WMAP9}%
  \BibitemOpen
  \bibfield{author}{%
  \bibinfo {author} {\bibfnamefont{G.}~\bibnamefont{{Hinshaw}}}, \bibinfo
  {author} {\bibfnamefont{D.}~\bibnamefont{{Larson}}}, \bibinfo {author}
  {\bibfnamefont{E.}~\bibnamefont{{Komatsu}}}, \bibinfo {author}
  {\bibfnamefont{D.~N.}\ \bibnamefont{{Spergel}}}, \emph{et~al.},\ }%
  \bibfield{journal}{%
  \Doi{10.1088/0067-0049/208/2/19}{\bibinfo {journal} {ApJSÊ}}\ }%
  \textbf{\bibinfo {volume} {208}},\ \bibinfo {eid} {19} (\bibinfo {year}
  {2013}),\ \Eprint{http://arxiv.org/abs/1212.5226}{arXiv:1212.5226
  [astro-ph.CO]}%
  \bibAnnoteFile{NoStop}{WMAP9}%
\bibitem{PLANCK_PS}%
  \BibitemOpen
  \bibfield{author}{%
  \bibinfo {author} {\bibnamefont{{Planck collaboration}}}}%
   (\bibinfo {year} {2013}),\
  \Eprint{http://arxiv.org/abs/1303.5075}{arXiv:1303.5075 [astro-ph.CO]}%
  \bibAnnoteFile{NoStop}{PLANCK_PS}%
\bibitem{KKS}%
  \BibitemOpen
  \bibfield{author}{%
  \bibinfo {author} {\bibfnamefont{M.}~\bibnamefont{{Kamionkowski}}}, \bibinfo
  {author} {\bibfnamefont{A.}~\bibnamefont{{Kosowsky}}},\ and\ \bibinfo
  {author} {\bibfnamefont{A.}~\bibnamefont{{Stebbins}}},\ }%
  \bibfield{journal}{%
  \Doi{10.1103/PhysRevD.55.7368}{\bibinfo {journal} {\prd}}\ }%
  \textbf{\bibinfo {volume} {55}},\ \bibinfo {pages} {7368} (\bibinfo {year}
  {1997}),\ \Eprint{http://arxiv.org/abs/astro-ph/9611125}{astro-ph/9611125}%
  \bibAnnoteFile{NoStop}{KKS}%
\bibitem{CLASS}%
  \BibitemOpen
  \bibfield{author}{%
  \bibinfo {author} {\bibfnamefont{D.}~\bibnamefont{{Blas}}}, \bibinfo {author}
  {\bibfnamefont{J.}~\bibnamefont{{Lesgourgues}}},\ and\ \bibinfo {author}
  {\bibfnamefont{T.}~\bibnamefont{{Tram}}},\ }%
  \bibfield{journal}{%
  \Doi{10.1088/1475-7516/2011/07/034}{\bibinfo {journal} {JCAP}}\ }%
  \textbf{\bibinfo {volume} {7}},\ \bibinfo {eid} {034} (\bibinfo {year}
  {2011}),\ \Eprint{http://arxiv.org/abs/1104.2933}{arXiv:1104.2933
  [astro-ph.CO]}%
  \bibAnnoteFile{NoStop}{CLASS}%
\bibitem{HALOFIT}%
  \BibitemOpen
  \bibfield{author}{%
  \bibinfo {author} {\bibfnamefont{R.~E.}\ \bibnamefont{{Smith}}}, \bibinfo
  {author} {\bibfnamefont{J.~A.}\ \bibnamefont{{Peacock}}}, \bibinfo {author}
  {\bibfnamefont{A.}~\bibnamefont{{Jenkins}}}, \bibinfo {author}
  {\bibfnamefont{S.~D.~M.}\ \bibnamefont{{White}}}, \bibinfo {author}
  {\bibfnamefont{C.~S.}\ \bibnamefont{{Frenk}}}, \bibinfo {author}
  {\bibfnamefont{F.~R.}\ \bibnamefont{{Pearce}}}, \bibinfo {author}
  {\bibfnamefont{P.~A.}\ \bibnamefont{{Thomas}}}, \bibinfo {author}
  {\bibfnamefont{G.}~\bibnamefont{{Efstathiou}}},\ and\ \bibinfo {author}
  {\bibfnamefont{H.~M.~P.}\ \bibnamefont{{Couchman}}},\ }%
  \bibfield{journal}{%
  \Doi{10.1046/j.1365-8711.2003.06503.x}{\bibinfo {journal} {MNRAS}}\ }%
  \textbf{\bibinfo {volume} {341}},\ \bibinfo {pages} {1311} (\bibinfo {year}
  {2003}),\ \Eprint{http://arxiv.org/abs/astro-ph/0207664}{astro-ph/0207664}%
  \bibAnnoteFile{NoStop}{HALOFIT}%
\bibitem{PEAK_AKW}%
  \BibitemOpen
  \bibinfo {author} {\bibfnamefont{S.}~\bibnamefont{{Aiola}}}, \bibinfo
  {author} {\bibfnamefont{A.}~\bibnamefont{{Kosowsky}}},\ and\ \bibinfo
  {author} {\bibfnamefont{B.}~\bibnamefont{{Wang~-~In preparation}}}%
  \bibAnnoteFile{NoStop}{PEAK_AKW}%
\bibitem{HEALPix}%
  \BibitemOpen
\bibfield{author}{%
    }%
  \bibfield{author}{%
  \bibinfo {author} {\bibfnamefont{K.~M.}\ \bibnamefont{{G{\'o}rski}}},
  \bibinfo {author} {\bibfnamefont{E.}~\bibnamefont{{Hivon}}}, \bibinfo
  {author} {\bibfnamefont{A.~J.}\ \bibnamefont{{Banday}}}, \bibinfo {author}
  {\bibfnamefont{B.~D.}\ \bibnamefont{{Wandelt}}}, \bibinfo {author}
  {\bibfnamefont{F.~K.}\ \bibnamefont{{Hansen}}}, \bibinfo {author}
  {\bibfnamefont{M.}~\bibnamefont{{Reinecke}}},\ and\ \bibinfo {author}
  {\bibfnamefont{M.}~\bibnamefont{{Bartelmann}}},\ }%
  \bibfield{journal}{%
  \Doi{10.1086/427976}{\bibinfo {journal} {\apj}}\ }%
  \textbf{\bibinfo {volume} {622}},\ \bibinfo {pages} {759} (\bibinfo {year}
  {2005}),\ \Eprint{http://arxiv.org/abs/astro-ph/0409513}{astro-ph/0409513}%
  \bibAnnoteFile{NoStop}{HEALPix}%
\bibitem{PLANCK_CS}%
  \BibitemOpen
  \bibfield{author}{%
  \bibinfo {author} {\bibnamefont{{Planck Collaboration}}}}%
   (\bibinfo {year} {2013}),\
  \Eprint{http://arxiv.org/abs/1303.5072}{arXiv:1303.5072 [astro-ph.CO]}%
  \bibAnnoteFile{NoStop}{PLANCK_CS}%
\bibitem{Bobin}%
  \BibitemOpen
  \bibfield{author}{%
  \bibinfo {author} {\bibfnamefont{J.}~\bibnamefont{{Bobin}}}, \bibinfo
  {author} {\bibfnamefont{F.}~\bibnamefont{{Sureau}}}, \bibinfo {author}
  {\bibfnamefont{J.-L.}\ \bibnamefont{{Starck}}}, \bibinfo {author}
  {\bibfnamefont{A.}~\bibnamefont{{Rassat}}},\ and\ \bibinfo {author}
  {\bibfnamefont{P.}~\bibnamefont{{Paykari}}},\ }%
  \bibfield{journal}{%
  \Doi{10.1051/0004-6361/201322372}{\bibinfo {journal} {A\&A Ê}}\ }%
  \textbf{\bibinfo {volume} {563}},\ \bibinfo {eid} {A105} (\bibinfo {year}
  {2014}),\ \Eprint{http://arxiv.org/abs/1401.6016}{arXiv:1401.6016
  [astro-ph.CO]}%
  \bibAnnoteFile{NoStop}{Bobin}%
\bibitem{PLANCK_GEN}%
  \BibitemOpen
  \bibfield{author}{%
  \bibinfo {author} {\bibnamefont{{Planck Collaboration}}}}%
   (\bibinfo {year} {2013}),\
  \Eprint{http://arxiv.org/abs/1303.5062}{arXiv:1303.5062 [astro-ph.CO]}%
  \bibAnnoteFile{NoStop}{PLANCK_GEN}%
\bibitem{ilic2013}%
  \BibitemOpen
  \bibfield{author}{%
  \bibinfo {author} {\bibfnamefont{S.}~\bibnamefont{{Ili{\'c}}}}, \bibinfo
  {author} {\bibfnamefont{M.}~\bibnamefont{{Langer}}},\ and\ \bibinfo {author}
  {\bibfnamefont{M.}~\bibnamefont{{Douspis}}},\ }%
  \bibfield{journal}{%
  \Doi{10.1051/0004-6361/201321150}{\bibinfo {journal} {A\&A Ê}}\ }%
  \textbf{\bibinfo {volume} {556}},\ \bibinfo {eid} {A51} (\bibinfo {year}
  {2013}),\ \Eprint{http://arxiv.org/abs/1301.5849}{arXiv:1301.5849
  [astro-ph.CO]}%
  \bibAnnoteFile{NoStop}{ilic2013}%
\bibitem{skymapper}%
  \BibitemOpen
  \bibfield{author}{%
  \bibinfo {author} {\bibfnamefont{S.~C.}\ \bibnamefont{{Keller}}}, \bibinfo
  {author} {\bibfnamefont{B.~P.}\ \bibnamefont{{Schmidt}}}, \emph{et~al.},\ }%
  \bibfield{journal}{%
  \Doi{10.1071/AS07001}{\bibinfo {journal} {PASA}}\ }%
  \textbf{\bibinfo {volume} {24}},\ \bibinfo {pages} {1} (\bibinfo {year}
  {2007}),\ \Eprint{http://arxiv.org/abs/astro-ph/0702511}{astro-ph/0702511}%
  \bibAnnoteFile{NoStop}{skymapper}%
\bibitem{lsst}%
  \BibitemOpen
  \bibfield{author}{%
  \bibinfo {author} {\bibnamefont{{LSST Science Collaboration}}}}%
   (\bibinfo {year} {2009}),\
  \Eprint{http://arxiv.org/abs/0912.0201}{arXiv:0912.0201 [astro-ph.IM]}%
  \bibAnnoteFile{NoStop}{lsst}%
\end{thebibliography}%
\end{document}